\documentclass[twocolumn,runningheads]{svjour2}
\smartqed  
\usepackage{graphicx}
\journalname{Astrophysics and Space Science}
\begin{document}

\title{Population statistics study of radio and gamma-ray pulsars in the Galactic plane
}


\author{Peter L. Gonthier         \and
        Sarah A. Story		\and
        Brian D. Clow		\and
        Alice K. Harding
}


\institute{P.L. Gonthier \and S.A. Story \and B.D. Clow  \at
              Hope College, Department of Physics, 27 Graves Place, Holland, MI 49424 \\
              Tel.: 616-395-7142\\
              Fax: 616-395-7123\\
              \email{gonthier@hope.edu,sarah.story@hope.edu, \\ brian.clow@hope.edu}
           \and
           A. Harding \at
              NASA Goddard Space Flight Center, Laboratory for High Energy Astrophysics, Greenbelt, MD 20771 \\
              Tel.: 301-286-7824\\
              Fax: 301-286-1682\\
              \email{harding@twinkie.gsfc.nasa.gov}           
           \and
}

\date{Received: date / Accepted: date}

\maketitle

\begin{abstract}
We present results of our pulsar population synthesis of ordinary
isolated and millisecond pulsars in the Galactic plane.
Over the past several years, a
program has been developed to simulate pulsar birth, evolution and
emission using Monte Carlo techniques. We have added to the program the
capability to simulate millisecond pulsars, which are old, recycled
pulsars with extremely short periods.  We model the spatial distribution
of the simulated pulsars by assuming that they start with a random kick
velocity and then evolve through the Galactic potential.  We use a polar
cap/slot gap model for $\gamma$-ray emission from both millisecond and
ordinary pulsars.  From our studies of radio pulsars that have clearly identifiable
core and cone components, in which we fit the polarization sweep as well
as the pulse profiles in order to constrain the viewing geometry, we
develop a model describing the ratio of radio core-to-cone peak fluxes. 
In this model, short period pulsars are more cone-dominated than in our
previous studies.  We present the preliminary results of our recent
study and the implications for observing these pulsars with GLAST and
AGILE.
\end{abstract}

\section{Introduction}
\label{intro}

In the very near future, the $\gamma$-ray telescopes AGILE and GLAST
will be launched.  It is expected that the number of
identified $\gamma$-ray pulsars will greatly increase.  
The $\gamma$-ray pulsar Geminga appears to be radio silent or at
least very weak.  
It is not clear whether its radio quiet nature is caused by a misalignment
of $\gamma$-ray and radio beams due to viewing geometry, or by intrinsically 
weak radio emission.
In the past, polar cap models \cite{Dau1996} located the
$\gamma$-ray emission along the same open field lines as radio emission
at similar altitudes.  In such a model, the $\gamma$-ray and radio beams
are coaxial with a large overlap, predicting a larger number of
radio-loud than radio-quiet $\gamma$-ray pulsars. On the other hand, the
$\gamma$-ray emission region in the outer gap models
\cite{Rom1995}\cite{Chen2000} is located near the light cylinder where the
$\gamma$-ray beam is at a large angle relative to the radio beam that leads the
viewer to observe high-energy and
radio emission originating from opposite poles.  The outer gap model
predicts a much larger number of radio-quiet than radio-loud
$\gamma$-ray pulsars. Thus the ratio of radio-loud to radio-quiet
$\gamma$-ray pulsars could serve as a discriminating signature to
distinguish between competing models. 

\section{Simulation - Assumptions}\label{sec:1} 
The present simulations are extensions of
previous works by Gonthier et al. (\cite{Gon2002}, \cite{Gon2004} \&
\cite{Gon2006}) that include both normal, isolated and millisecond
pulsars from the Galactic disk, and the emission
of $\gamma$-rays within the slot gap/polar cap model. 
Two years ago at the Hong Kong meeting \cite{Gon2005}, we presented
simulations of pulsars from the Gould Belt accounting for less than 30\%
of the EGRET unidentified $\gamma$-ray sources that were previously
correlated to the location of the Gould Belt. Recent reassessment of the
diffuse $\gamma$-ray background by Casandjian \& Grenier \cite{Cas2006}
suggests that the Gould Belt is no longer such a significant source of
EGRET unidentified $\gamma$-ray sources.  As a result, we do not
simulate pulsars from the Gould Belt in this study.

For both normal and millisecond (ms) pulsars, we assume a  birth
location in the Galactic disk as given by Paczy\'{n}ski \cite{Pac1990}. 
We evolve the neutron stars from their birth location to the present in
the Galactic potential defined by Dehnen \& Binney \cite{Dehn1998}.  For
normal pulsars, we assume a supernova kick velocity distribution of
Hobbs et al. \cite{Hobb2005}, a uniform initial period distribution from
0 to 500 ms, initial magnetic field distributions with a decay constant
as given in Gonthier et al. \cite{Gon2004} and a uniform birth rate back
to 1 Gyr.

For ms pulsars, we begin the evolution by using the
magnetic field and supernova kick velocity distributions of Cordes \&
Chernoff \cite{Cord1997}. Recent studies of low-mass X-ray binary
systems (LMXBs) have been able to determine the spins of the accreting
neutron stars allowing for an estimate of their magnetic fields. Lamb \&
Yu \cite{Lamb2005} conclude that the properties of these LMXBs are
consistent, if they have magnetic fields between $3\times 10^7$ G and
$3\times 10^8$ G and accretion rates ranging from the Eddington critical
rate $\dot{M}_E$ to $3\times 10^{-3}\dot{M}_E$. These different
accretion rates result in different birth lines in the $P-\dot{P}$
diagram (see figure 4 in Lamb \& Yu \cite{Lamb2005}).  We have included
an approximate procedure to take into account this distribution of
accretion rates by dithering the intercept of the birth line described
as

 \begin{equation}
 \centering
 \log(\dot{P})={4\over 3}\log(P)-14.9.
 \end{equation}

While we tried a Gaussian distribution of the birth line, we obtained
better agreement with a uniform distribution of birth lines between
$\dot{M}_E$ to $3\times 10^{-3}\dot{M}_E$\cite{Lamb2005}.  We assume a
uniform birth rate for ms pulsars back to 12 Gyr. We explore various
power laws of the magnetic field distribution getting better agreement
with $n(B)\propto B^{-1}$ and with a $B_{min}=2\times 10^{8} G$.  Given
the magnetic field, a selected birth line and the pulsar age, we obtain
the present period and period derivative.

We use a larger Galactic scale height of 200 pc for ms pulsars, compared
to the one used for normal pulsars of 75 pc \cite{Pac1990}.  Since the
supernova kick velocities of ms pulsars are much smaller than those of
normal pulsars, most ms pulsars remain bound to the Galaxy, oscillating
in and out of the Galactic plane with time.  We evolve a large number of
neutron stars generated in our Monte Carlo simulation to determine the
equilibrium spatial distribution of ms pulsars, which we then use in
subsequent simulations.  A scale height of 410 pc of the evolved ms
pulsars is in good agreement with a scale height of 500 pc (exponential
scale) of Cordes \& Chernoff \cite{Cord1997} and with a scale height of
410 pc of LMXBs \cite{Grim2002}.

For the radio luminosity, we assume that radio pulsars are standard candles and 
follow the prescription of Arzoumanian, Chernoff \& Cordes \cite{Arzo2002} (ACC)
given by the expression 
\begin{equation}
\centering
L = 2.1\times 10^{12}P^{-1.3}\dot{P}^{0.4} {\rm mJy\cdot kpc^2\cdot MHz}.
\end{equation}

However, to obtain good agreement between the measured and simulated flux
and distance distributions, we needed to reduce the luminosity by a factor
of 73 and 200 for normal pulsars and ms pulsars, respectively. We follow
the geometric model of the core beam in ACC \cite{Arzo2002} and the cone
model of Kijak \& Gil \cite{Kija1998} described in this meeting by
Harding, Grenier, \& Gonthier \cite{Hard2006}.  In the study of ACC, the
ratio of the core-to-cone peak fluxes had a $P^{-1}$ dependence,
resulting in a dominance of the core component for short period pulsars.
As discussed by Harding et al. \cite{Hard2006}, we find compelling
evidence that short period pulsars are more cone dominated than in the
ACC model and have adopted a different model to characterize this
relationship (see \cite{Hard2006}).

The ``detection'' of our simulated radio pulsars is accomplished using the
characteristics of ten radio surveys, six of which are at a frequency
around 400 MHz - Arecibo 3, Arecibo 2, Greenbank 2, Greenbank 3, Molongo
2 and Parkes 2, and four surveys are around 1400 MHz - Parkes 1, Jodrell
Bank 2, Parkes Multibeam, and we have recently added the Swinburne
Intermediate Latitude survey at 1400 MHz.  These surveys provide us with 1208 normal
and 50 ms pulsars to which we normalize our simulation.  We adjust the
radio luminosity using only the Parkes MB pulsar survey in order to get
a neutron star birth rate of about 2 per century.  This survey has over
800 detected pulsars, and we have the best description of the minimum radio 
flux $S_{\rm min}$ of the ten surveys.  Normalization to 1208 radio pulsars
seen by the group of 10 surveys is then the only overall adjustment
made, thereby, allowing the prediction of birth rates, radio-loud and
radio-quiet $\gamma$-ray pulsars detected by the instruments EGRET, AGILE
and GLAST. Simulated neutron stars whose radio flux is below the survey
threshold, $S_{\rm min}$, are assumed to be radio-quiet.

The simulation of the $\gamma$-ray emission occurs over two regions in the
$P-\dot{P}$ diagram separated by the curvature radiation pair death line
(CRPDL), below which the curvature radiation $\gamma$ rays no longer produce
electron-positron pairs.  Above this death line, emission originates from 
low altitude pair cascades on the inner edge of the slot
gap \cite{Gon2004}, as well as from primaries accelerating in the slot gap at 
high altitude forming a caustic component \cite{Hard2006}.
Below this death line, the slot gap dissolves due to the less effective
screening of the electric field, leading
to extended emission over the entire polar cap.  The $\gamma$-ray emission
above CRPDL is most important for normal pulsars while emission below the CRPDL  
is the dominant $\gamma$-ray emission mechanism for ms pulsars.

The simulated $\gamma$-ray flux is compared to the all sky threshold maps
for EGRET, AGILE and GLAST.  We include the revised EGRET map that
includes the dark clouds (see \cite{Cas2006}) and the revised GLAST map (after DC2)
without the dark clouds (Grenier private communication).  The all sky map for
AGILE (Pellizzoni private communication) has not been recently updated.

\section{Results}\label{sec:2}

\begin{figure*}
\centering
  \includegraphics[width=1.0\textwidth]{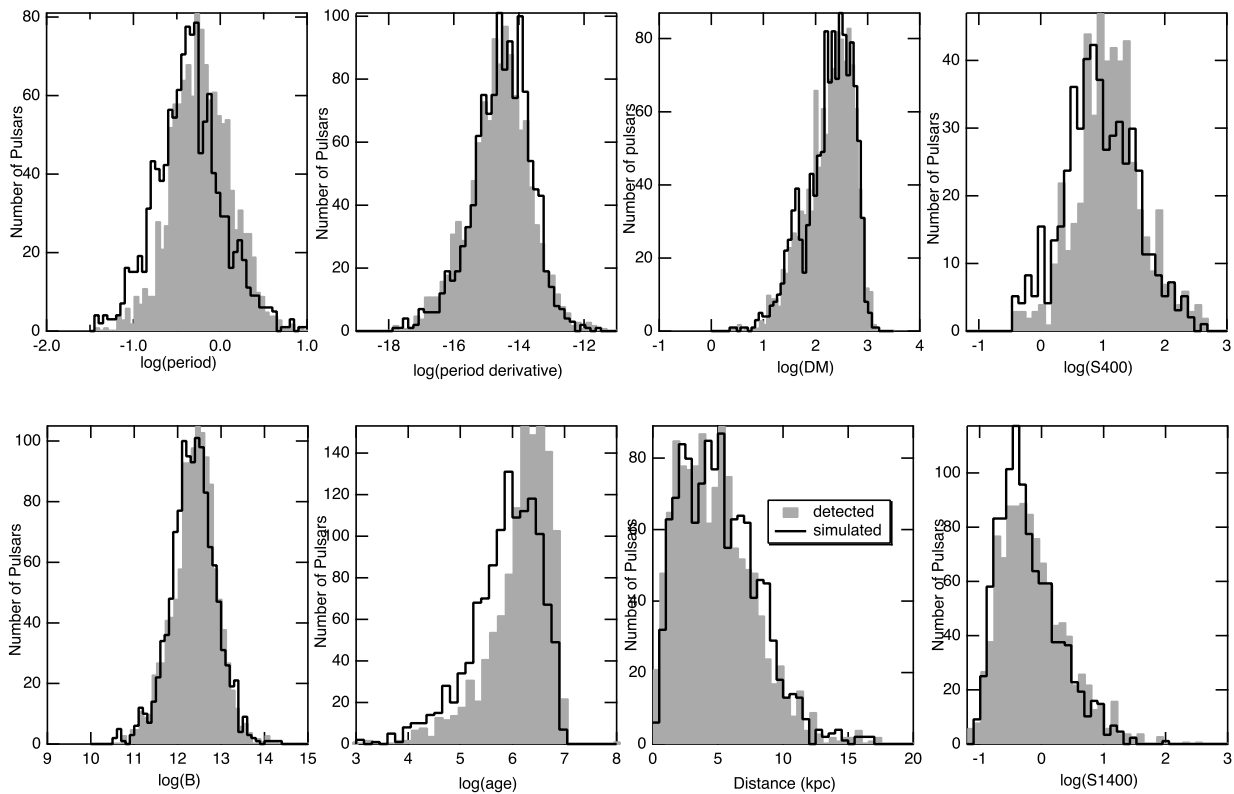}
\caption{Distributions of various  characteristics of normal pulsars indicated as detected pulsars
(shaded histograms) and simulated (unshaded histograms) pulsars from the Galactic plane.}
\label{fig:1}       
\end{figure*}

In Figure \ref{fig:1}, we compare simulated distributions (unshaded histograms) of
various characteristics of normal pulsars with those detected (shaded
histograms) by the select group of 10 radio surveys.
For the flux distributions at 400 and 1400 MHz, we have used the values
given in the ATNF pulsar catalog for the detected distributions.  In the
simulation, we assign a flux at 400 MHz if the pulsar is detected by one
of the low frequency surveys in our group and likewise for the fluxes at
1400 MHz. As can be seen in the comparisons of these distributions, the
model simulation over-predicts the number of young pulsars with short
periods.  The distance distribution is now much improved from our
previous results \cite{Gon2006} mainly due to the inclusion of the
actual width of the simulated pulse profile of each pulsar in the
calculation of the survey flux threshold, $S_{\rm min}$.  Overall we
generally see good agreement between the simulation and detected
distributions with a predicted birth rate of 1.8 normal pulsars per century.

The simulation again overestimates the number of young and distant ms 
pulsars as shown in Figure 2.
\begin{figure*}
\centering
  \includegraphics[width=1.0\textwidth]{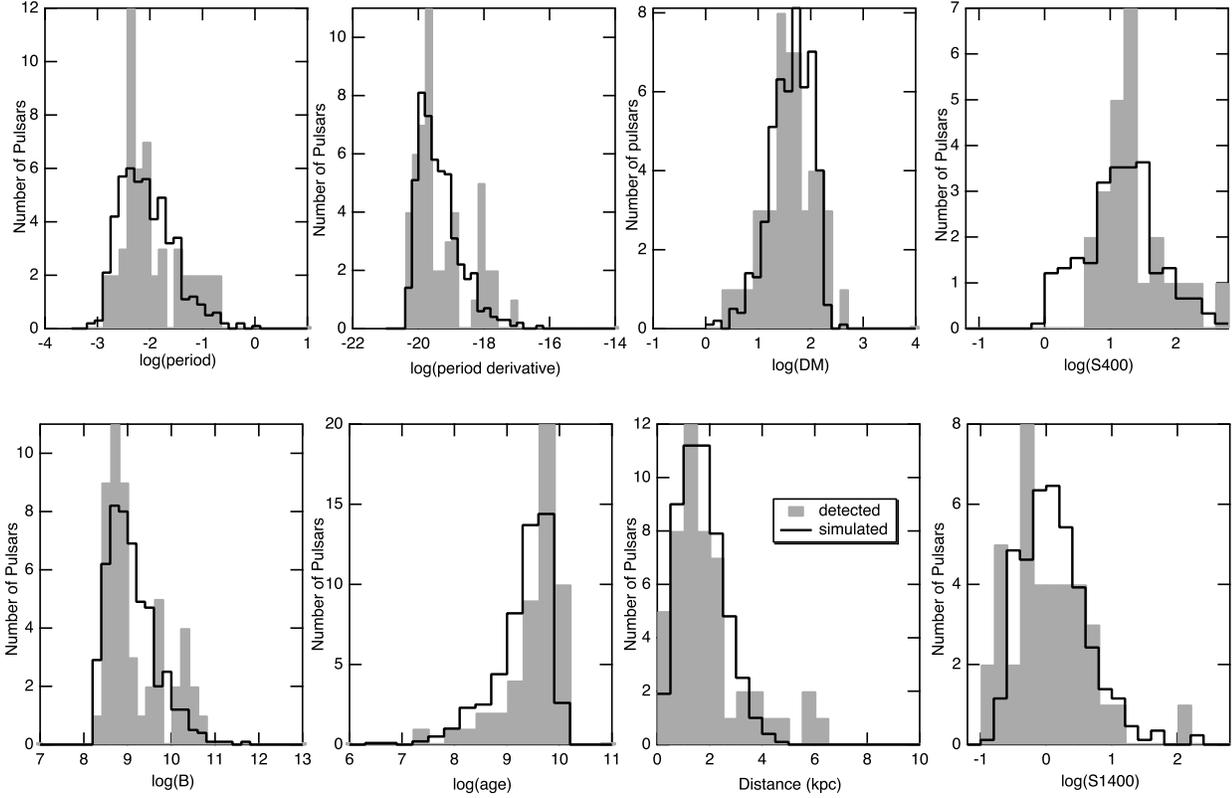}
\caption{Distributions of various characteristics of millisecond pulsars indicated as detected
(shaded histograms) and simulated (open histograms) pulsars from the Galactic plane.}
\label{fig:2}       
\end{figure*}
Evident in the histograms is the limitation of a sample of 50 ms pulsars
detected by the select group of ten radio surveys.  We simulate 500 ms pulsars and
then normalize to the detected 50 in order to obtain smoother histograms for
the simulated pulsars.  The simulation again
overestimates young ms pulsars that are more distance than those
detected.  The 400 MHz surveys in the simulation seem to be a bit too
sensitive.  The predicted birth rate is $6.8\times 10^{-4}$ per century,
which is very close to the birth rates estimated by Lorimer
\cite{Lorm2005} of $2.9\times 10^{-4}$ per century and by Kiel \& Hurley
\cite{Kiel2006} of $6.5\times 10^{-4}$ per century of neutron stars from
LMXBs.
%
%

In Figure 3, we present the Aitoff plots and $P-\dot{P}$ diagrams for detected (left)
and simulated (right) normal (dots) and MS (crosses) pulsars.
\begin{figure*}
\centering
  \includegraphics[width=1.0\textwidth]{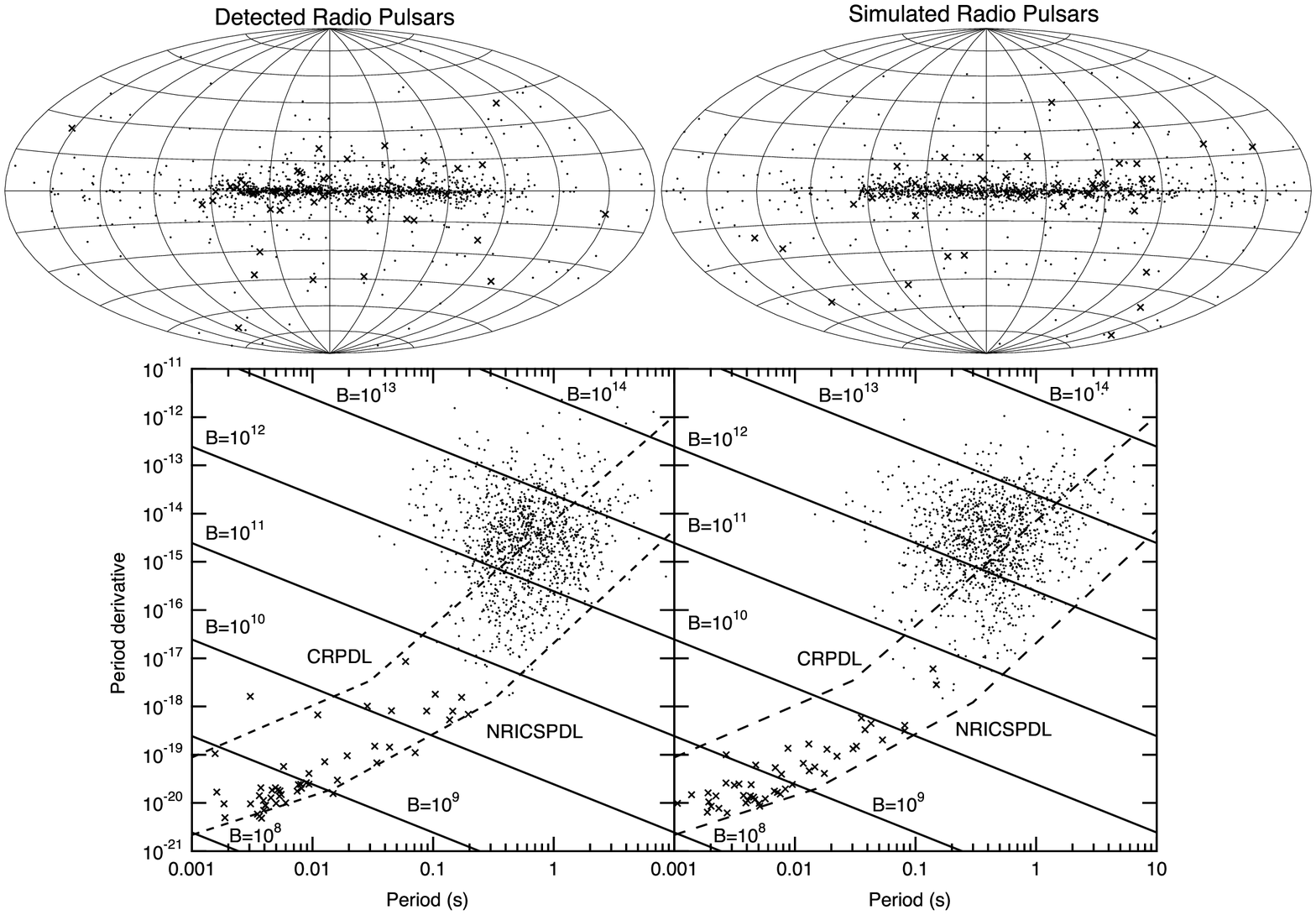}
\caption{Aitoff plots of normal (dots) and millisecond (crosses) pulsars
detected (left) by the select group of ten radio surveys and simulated
(right).  
Lower plots are the $P-\dot{P}$ diagrams of normal (dots) and
millisecond (crosses) pulsars detected (left) and simulated (right). 
Dashed lines represent the pair death lines for curvature radiation
(CRPDL) and for nonresonant inverse Compton scattering (NRICSPDL). Solid
lines represent the traditional magnetic surface field strength,
assuming a constant dipole spin-down field.}
\label{fig:3}       
\end{figure*}
Both simulated normal and ms pulsar distributions in the Galaxy are
similar to the distributions of those detected.  As ms pulsars are closer
than normal pulsars, they appear with a larger out-of-plane
distribution.  We are not quite able to reproduce the``bunching" of ms
pulsars with periods between 2 and 10 ms.  However, the uniform
distribution of the birth line as suggested by Lamb \& Yu
\cite{Lamb2005} does seem to be important in order to get fairly good
agreement in the $P-\dot{P}$ diagram with detected ms pulsars.  The
distribution of detected normal pulsars in the $P-\dot{P}$ diagram is more or
less reproduced by the simulation, but the simulated distribution is too
broad in period with too few high-field pulsars.  We do find that
magnetic field decay with a decay constant of 2.8 Myr is necessary for
normal pulsars in order to reproduce the detected distribution.  We did not
incorporate field decay in the simulation of ms pulsars.

\begin{table*}[t]
\caption{Simulated radio-loud and radio-quiet $\gamma$-ray pulsar statistics}
\centering
\label{tab:1}       
\begin{tabular}{lcccc}
\hline\noalign{\smallskip}
 & \multicolumn{2}{c}{Normal Pulsars} & \multicolumn{2}{c}{Millisecond Pulsars}  \\[3pt]
 & \multicolumn{2}{c}{low + high altitude} & \multicolumn{2}{c}{ }  \\[3pt]
\tableheadseprule\noalign{\smallskip} \hline
 Instrument & Radio-Loud & Radio-Quiet & Radio-Loud & Radio-Quiet \\ \hline
EGRET Detected & 6 & 1 & 1 & 0\\
EGRET Simulated & 25+2 & 10+22 & 1 & 2 \\
AGILE Simulated & 51+1 & 16+53 & 2 & 3 \\
GLAST Simulated & 94+6 & 51+100 & 4 & 10 \\ \hline
\noalign{\smallskip}\hline
\end{tabular}
\end{table*}

In Table 1, we present the simulated normal and ms $\gamma$-ray pulsar
statistics for radio-loud and radio-quiet $\gamma$-ray pulsars for
various instruments indicated as well as the ones detected by EGRET. 
For normal pulsars, we show separately the number of $\gamma$-ray
pulsars having detected emission from the low altitude and high altitude slot gap, as
both of these radiation mechanisms contribute above the curvature
radiation pair death line.  As expected the low altitude slot gap beam
is more co-axial with the radio beam geometry and results in a greater
number of radio-loud $\gamma$-ray pulsars.  On the other hand, the high
altitude emission of the slot gap occurs along the last open field lines
all the way out to the light cylinder, at which point the $\gamma$-ray
beam makes a larger angle to the radio beam and more radio-quiet
$\gamma$-ray pulsars are detected.
In this respect, the high altitude emission of the slot gap
resembles the correlations of radio and $\gamma$-ray beams of outer gap
emission.

The ms $\gamma$-ray pulsars in the $P-\dot{P}$ diagram appear below the
curvature radiation pair death line and, therefore, their $\gamma$-ray
emission arises from the emission over the entire polar cap region as
the slot gap geometry disappears near the curvature death line.

\section{Conclusions}\label{sec:3}
We present the preliminary results of a population synthesis study of
both normal and millisecond pulsars.  We include a radio beam geometry
that includes a single core and a single cone beam with a new
dependence on the core-to-cone peak fluxes. In this new model, as 
discussed in this conference by Harding et al.
\cite{Hard2006}, short period pulsars are more cone dominated than in our 
previous studies that follow the ACC
model \cite{Arzo2002}. This relationship is especially important for
millisecond pulsars.  We use the same radio model for both normal and
millisecond pulsars.  We describe the radio luminosity using the
prescription of ACC (Equation 2), but with luminosity decreased by
a factor of 73  and 200 for normal and ms pulsars, respectively, to achieve 
reasonable neutron star birth rates, flux and distance distributions.  The same
set of parameters are then used in the simulation of normal and
millisecond pulsars.  The simulations are run until the same number of
normal or millisecond radio pulsars are simulated as detected by a
select group of ten radio surveys providing an overall normalization of
the simulation.  With the same set of parameters for the radio beam
geometry and luminosity, we achieve very reasonable agreement with the
detected distributions of various pulsar characteristics for both normal
and millisecond pulsars. This conclusion is quite remarkable.

However, we expect the radio beam geometry of millisecond pulsars to be
significantly different, especially for pulsars with periods less than
100 ms, due to the special relativistic effects of aberration, time
delays and the sweepback of the magnetic field.  These effects and
their contributions are discussed in this meeting by Harding et al.
\cite{Hard2006}.  We hope to soon include this component to the radio
emission in our Monte Carlo code to more adequately describe millisecond
pulsars.

In previous studies, we only included low altitude $\gamma$-ray emission
from the slot gap/polar cap model.  Both curvature and
synchrotron radiation contribute along the last open field line about
three stellar radii above the surface resulting in a conical beam
symmetric about the magnetic axis.  As a result, the $\gamma$-ray beam is
strongly correlated to the core and cone radio beams assumed in our
simulations.  Such a correlation is reflected in the larger number of
radio-loud than radio-quiet $\gamma$-ray pulsars with a ratio of
ratio-loud to radio-quiet of 1.8. As discussed by Harding et al.
\cite{Hard2006} at this meeting, the electric field parallel to the
magnetic field along the very last open field lines is not screened, forming a
especially narrow slot gap in the case of short period pulsars. The $\gamma$-ray beams
at high altitude are concentrated (caustic) along the last open field
lines with the beam being at large angle to the radio emission.  High
altitude $\gamma$-ray emission then leads to a larger number of
radio-quiet $\gamma$-ray pulsars being simulated by this mechanism with
a ratio of radio-loud to radio-quiet of about 0.06 for this component as
indicated in Table 1. The viewing geometry defining the impact angle
will determine the respective contributions of the low and high altitude
emission.  These components will have different signatures in their
pulse profiles.

EGRET saw pulsed emission from only one $\gamma$-ray ms pulsar,
J0218+4342. Though not very optimistic, the simulations predict that
GLAST should detect of the order of 10 ms pulsars as point sources.  Of
the 10 radio-quiet $\gamma$-ray ms pulsars only 1 is expected to be detected
through blind searches.

\begin{sloppypar}
Our study of normal, isolated pulsars indicates that GLAST will be able
to detect about 100 radio-loud and 151 radio-quiet $\gamma$-ray pulsars.
These numbers are significantly lower than presented in Gonthier et
al.\cite{Gon2004}, primarily because the new GLAST point-source
detection threshold is higher after the great detailed study of the
GLAST second data challenge (DC2).  The challenge for GLAST will be to
detect the radio-quiet $\gamma$-ray pulsars through its ability to
perform blind searches with those pulsars whose $\gamma$-ray fluxes are
higher than $10^{-7}\ {\rm photons/({cm^2\cdot s}})$ (Grenier private
communication). Out of the 151 radio-quiet $\gamma$-ray pulsars
GLAST should be able to detect pulsations in 50 pulsars.  The
simulation modeling the high and low altitude $\gamma$-ray emission
within the theoretical framework of the polar cap model suggests that
the expected ratio of radio-loud to radio-quiet $\gamma$-ray pulsars is about 
1.5, while the outer gap models predict many more
radio-quiet $\gamma$-ray pulsars and a ratio of about 0.1 radio-loud to
radio-quiet $\gamma$-ray pulsars is expected \cite{Jian2006}.  Note that
the outer gap ratio is similar to the ratio for the high altitude
emission alone. Further detailed studies of the correlations between the
light curves of radio and $\gamma$-ray pulses may provide the best
signature to distinguish between the outer gap and polar cap models.
\end{sloppypar}

\begin{acknowledgements}
We express our gratitude for the generous support of the Michigan Space Grant Consortium, 
of Research Corporation (CC5813), of the National Science Foundation (REU and AST-0307365) 
and the NASA Astrophysics Theory Program.
\end{acknowledgements}



\end{document}